\documentclass[review]{elsarticle}

\usepackage{lineno,hyperref}
\modulolinenumbers[5]

\journal{Journal of \LaTeX\ Templates}
\usepackage{xspace}
\usepackage{pdfwidgets}
\usepackage{enumitem}
\usepackage{latexsym}
\usepackage{tikz}
\usepackage{pgfplots}
\usepackage{natbib}
\pgfplotsset{compat=newest}
\usepgfplotslibrary{groupplots}
\usetikzlibrary{pgfplots.groupplots}
\usepgfplotslibrary{colormaps}
\usetikzlibrary{pgfplots.colormaps}
\usetikzlibrary{calc}
\usepgfplotslibrary{external}
\tikzsetexternalprefix{figures/}
\usepackage[textsize=tiny]{todonotes}
\makeatletter
\renewcommand{\todo}[2][]{\tikzexternaldisable\@todo[#1]{#2}\tikzexternalenable}
\makeatother

\makeatletter
\def\bs{\expandafter\@gobble\string\\}
\def\lb{\expandafter\@gobble\string\{}
\def\rb{\expandafter\@gobble\string\}}
\def\@pdfauthor{C.V.Radhakrishnan}
\def\@pdftitle{elsarticle.cls -- A documentation}
\def\@pdfsubject{Document formatting with elsarticle.cls}
\def\@pdfkeywords{LaTeX, Elsevier Ltd, document class}

\DeclareRobustCommand{\LaTeX}{L\kern-.26em%
        {\sbox\z@ T%
         \vbox to\ht\z@{\hbox{\check@mathfonts
           \fontsize\sf@size\z@
           \math@fontsfalse\selectfont
          A\,}%
         \vss}%
        }%
     \kern-.15em%
    \TeX}
\makeatother

\usepackage{array}
\newcolumntype{P}[1]{>{\centering\arraybackslash}p{#1}}
\usepackage{amsmath}
\begin{document}

\begin{frontmatter}

\title{Examination of the Plasma located in PSI Ring Cyclotron}


\author[PSI]{N. J. Pogue\corref{mycorrespondingauthor}}
\ead{nathaniel.pogue@psi.ch}
\ead[url]{psi.ch}

\address[PSI]{Paul Scherrer Institut, Villigen, Switzerland}



\cortext[mycorrespondingauthor]{Corresponding author}


\author[PSI]{A. Adelmann\corref{mycorrespondingauthor}}
\author[PSI]{M. Schneider}
\author[PSI]{L. Stingelin\corref{mycorrespondingauthor}}

\begin{abstract}
A plasma has been observed inside the vacuum chamber of the PSI Ring Cyclotron. This ionized gas cloud maybe a substantial contributor to several interior components having reduced lifetimes. The plasma's generation has been directly linked to the voltage that is applied to the Flat Top Cavity through visual confirmation using CCD cameras. A spectrometer was used to correlate the plasma's intensity and ignition to the Flat Top Cavity voltage as well as to determine the composition of the plasma. This paper reports on the analysis of the plasma using spectroscopy. The spectrometer data was analyzed to determine the composition of the plasma and that the plasma intensity (luminosity) directly corresponds to the Flat Top voltage. The results showed that the plasma was comprised of elements consistent with the cyclotrons vacuum interior.
\end{abstract}

\begin{keyword}
\texttt{Flat Top \sep RF Cavity \sep Plasma \sep Spectroscopy \sep Cyclotron Design}
\MSC[2010] 00-01\sep  99-00
\end{keyword}

\end{frontmatter}

%
%
%

\section{Introduction}

\begin{figure}[]
\centering                                                           
   \includegraphics[scale=.95]{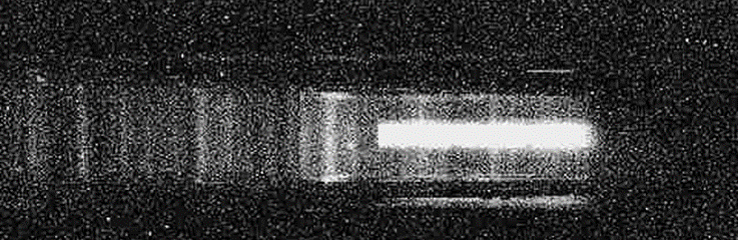}                               
  \caption{Image of the plasma located within the cyclotron vacuum space. This view is pointed from a flange on the outer wall of the FTC looking towards the neighboring sector magnet. A distributed glow appears from the left (near injection) to the right (near extraction), but it is unclear as to whether the plasma is completely contain within it. Though it maybe an optical illusion, the smaller band of light might be coming from the sector magnet behind the adjacent sector magnet. Further investigation is needed to determine the answer. Image courtesy of Richard Kan.}
  \label{plasma}                                                          
\end{figure}

The PSI HIPA Ring Cyclotron produces a 590 MeV, 2.4 mA, 1.4 MW proton beam. The Ring's power has continuously grown since 1974 through incremental upgrades to the system. It was first documented in 2009 \cite{Goetz}, that a plasma has been observed inside the vacuum space, shown in Figure \ref{plasma}. It has been conjectured that this plasma maybe adversely effecting several components inside the cyclotron by depositing metallic atoms upon critical insulating components such as the extraction septum. Additionally the plasma is thought to be significantly contributing to the number of beam trips in recent years. An excellent description of the problem and its dependencies can be found in D. Goetz et al. \cite{Goetz}. 

The length of time the plasma has existed inside the machine is debatable. However, one thing is clear, as the Flat Top Cavity's (FTC) voltage has been progressively increased over the years, the intensity of the plasma has also increased. Obtaining a fundamental understanding of the plasma is important as it will effect future developments and modification regarding the FTC.

The FTC is currently the most critical component in the system as it is the current limiting feature. The FTC increases the longitudinal acceptance of the Ring by decelerating the beam by 11\% of the combined gain in the four main accelerating cavities. However, the FTC has reached a maximum voltage due to leaking power and inability to keep the cavity at the correct frequency. A hydraulic system is the mechanism by which the frequency is kept in tune. The heating and thermal expansion of the cavity at higher voltages causes the cavity to deform to an extent for which the existing hydraulic system can not compensate. Thus understanding the plasma and its causes are of great importance.

\begin{figure}[]
\centering                                                           
   \includegraphics[scale=.36]{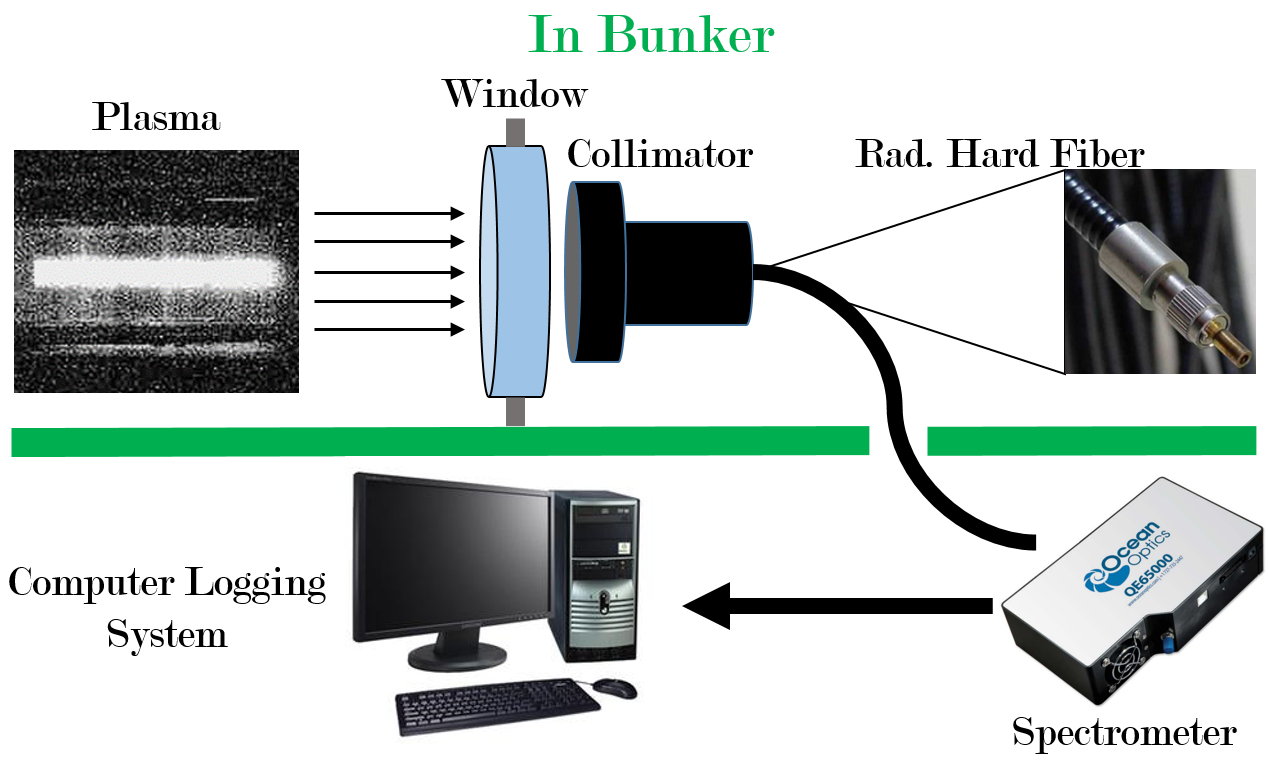}                               
  \caption{A schematic of the experimental setup. The plasma emits light that passes through a borosilicate window with a Indium Tin Oxide Coating (ITO). The light then passes through a collimator which allows the light to enter a radiation hard optical fiber. The fiber then channels the light out of the bunker to a spectrometer that quantifies the wavelength. The information is passed to the PSI logging system where the count number and spectrum can be analyzed.}
  \label{Ring}                                                          
\end{figure}

This paper expounds upon data from a spectrometer placed in a window near the FTC, shown in Figure \ref{Ring}, and other observations such as power measurements versus generated voltage in the cavity. The results provide a second method of confirmation that indeed a plasma does exist in the cyclotron, the intensity of the plasma is directly related to the voltage of the cavity, and that the composition of the plasma is constant with what would be expected.

\section{Experimental Setup}
An Ocean Optics Spectrometer QE6500, range 185-1100~nm, was the instrument chosen to view the spectra of the plasma. To provide a standard for calibration, an Ocean Optics DH-2000 light source was utilized to gauge the spectrometer’s response.  The spectrometer is comprised of several fragile components, and thus an optical fiber was used to transmit the light from the source to the spectrometer. The fiber of choice was the Mitsubishi Cable Industries FVSMA95P STU600F, 35 meter long, radiation hard fiber. The fiber must be radiation hard to prevent degradation of the signal over time. This is the only radiation hard fiber that could be found, and the length ensures that the spectrometer can be placed outside the cyclotron bunker.

\subsection{Setup and Calibration}
A blackbox was placed around the spectrometer, the rad-hard fiber, and a calibrated light source - Ocean Optics DH-2000. The data is clean and matches the output given by the source$’$s manufacturer. However, the signal below 350~nm is very low and cannot be calibrated – which does match the manufacturer$'$s specified light output profile for the source \citep{Ocean}. 

From previous experience, the window through which the fiber observes must also be radiation hard. The window used was a Schott BK7G radiation hard glass \cite{Schott1}. This window contains Cerium which keeps the window transparent even after long term radiation exposure. Studies by Hobson \cite{Schottrad} exposed the Schott BK7G glass to 1 MGy and only slight variations in transmittance occurred.

To prevent RF from the Flat Top cavity leaking out, the window is coated with an ITO coating. ITO (Indium Tin Oxide) is an industrial standard for this application. ITO has better performance compared to AZO (Aluminum Zinc Oxide) which is cheaper, but is more susceptible to damage. The ITO coating was only applied to one side of the window which reduces transmission losses.

The emitted light must travel through the coating and window to reach the fiber. Thus the opacity or transmission rate of these media must be taken into account. The coating deposited on the window is 20~$\Omega$/$\Box$, and therefore the ITO layer is several hundred nanometers thick. The Schott glass also has a wavelength dependent transmittance. Referring to Zhang \cite{ITO1} for the ITO transmittance, and Hobson \cite{Schottrad} for the Schott glass$'$s transmittance, any light emitted with wavelength below 350 nm will have no chance of reaching the spectrometer. As a result, the lack of calibration at these wavelengths is irrelevant as the coating and glass would not allow any light of these wavelengths to reach the spectrometer.

 \begin{figure}[h!]
\centering                                                           
   \includegraphics[scale=.4]{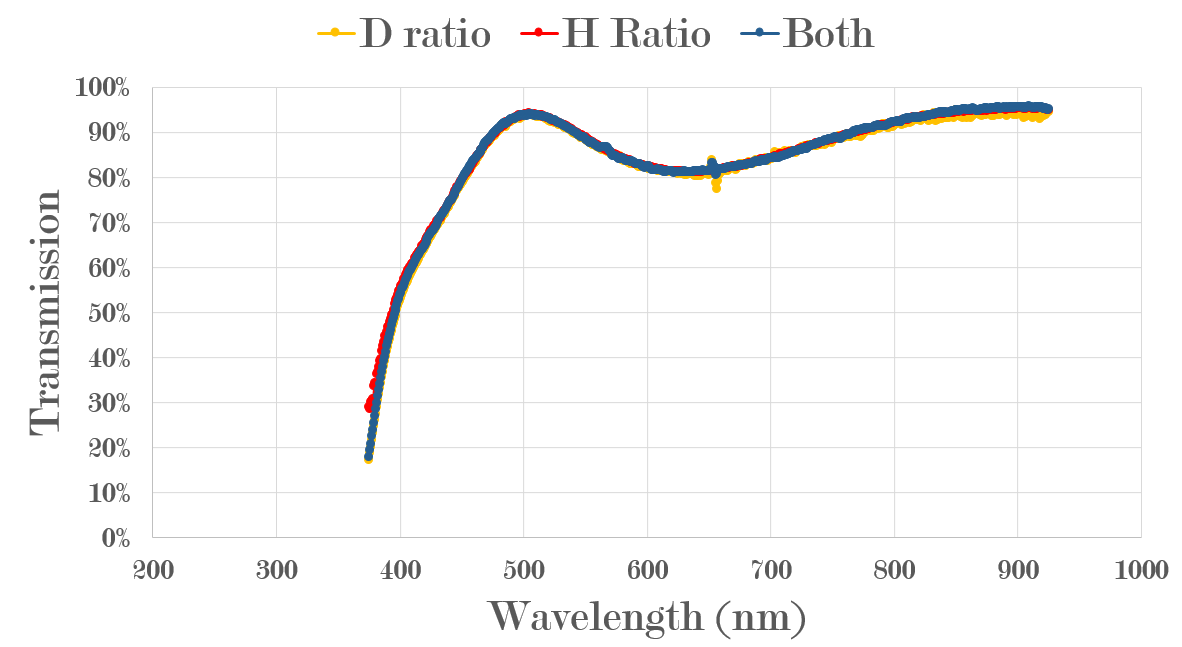}                               
  \caption{The percent transmission of light through the window compared to without is shown. The shape is similar to that of the ITO coating. The transmission curve was consistent for the three different option for light available. A curve was fitted that will be later used to produce the correct intensities.}
  \label{window-2}
\end{figure}

To determine the combined effect (window and coating), the coated window was placed between the calibration light source and fiber. The orientation of the window placed the film towards the light source, mimicking the set-up in the Ring bunker. Figure \ref{window-2} shows the percent transmission of light compared to the setup without the window. The shape is quite similar to that of an ITO coating on a borosilicate substrate \cite{Wiki}. The parameterized fit of the curve,

\begin{align}
y = &{}\  9.5200\cdot10^{-16}\cdot x^6 - 3.4584\cdot10^{-12}\cdot x^5 + 4.9690\cdot10^{-9}\cdot x^4  \\ 
&{} - 3.5371\cdot10^{-6}\cdot x^3 +1.2593\cdot10^{-3}\cdot x^2 - 1.8831\cdot10^{-1}\cdot x + 5.0160,\nonumber
\end{align} will be used to correctly scale the intensity of the spectrum when placed in the bunker. Thus the usable data range for the spectrometer is 350 to 925~nm.

\section{Measurements}
 
Using the EPICS control system for the HIPA complex, the current of the cyclotron and the RF voltage of the Flat Top and Main Cavities were plotted versus time. Several interesting time segments were identified. Background measurements were taken during periods where the beam and all the RF cavities were powered down. This background measurement, averaged over time, was subtracted from the other measurements to provide an accurate intensity relationship. For subsequent tests, a background period was sought near the time of the measurement. 

\subsection{Background}

The background measurements are the average of a minimum of 6 hour period. Most however measurements are over a 24 hour period. Though the number of maintenance periods are few, from the data acquired it can be clearly shown that the background negligibly varies. However, individual spectral lines vary slightly with each time average measurement. Hence, when determining the intensity, one must use a background measurement close in time to the spectral measurement.

Two other factors may have impact on the background. As the machine operates, the material will become progressively more radioactive. This activation is currently not seen. Either the increase in activation is very small in comparison to the current level, or the spectrometer is degrading in its performance. The spectrometer sees between 0-50 Rad at is location outside the bunker, and thus over long periods of time it might have a decrease in performance. Both reasons are under investigation.

Other parameters of interest were different levels of beam current and varied amounts of power inserted into the FTC, with and without the main cavities powered. Each parameter was varied independently to ensure causality could be achieved. The spectral intensity as a function of wavelength is described in the Experimental Setup section.

Subsection{Spectral Dependencies}
\subsection{Varying Current}
\begin{figure}[h!]
\centering                                                           
   \includegraphics[scale=.37]{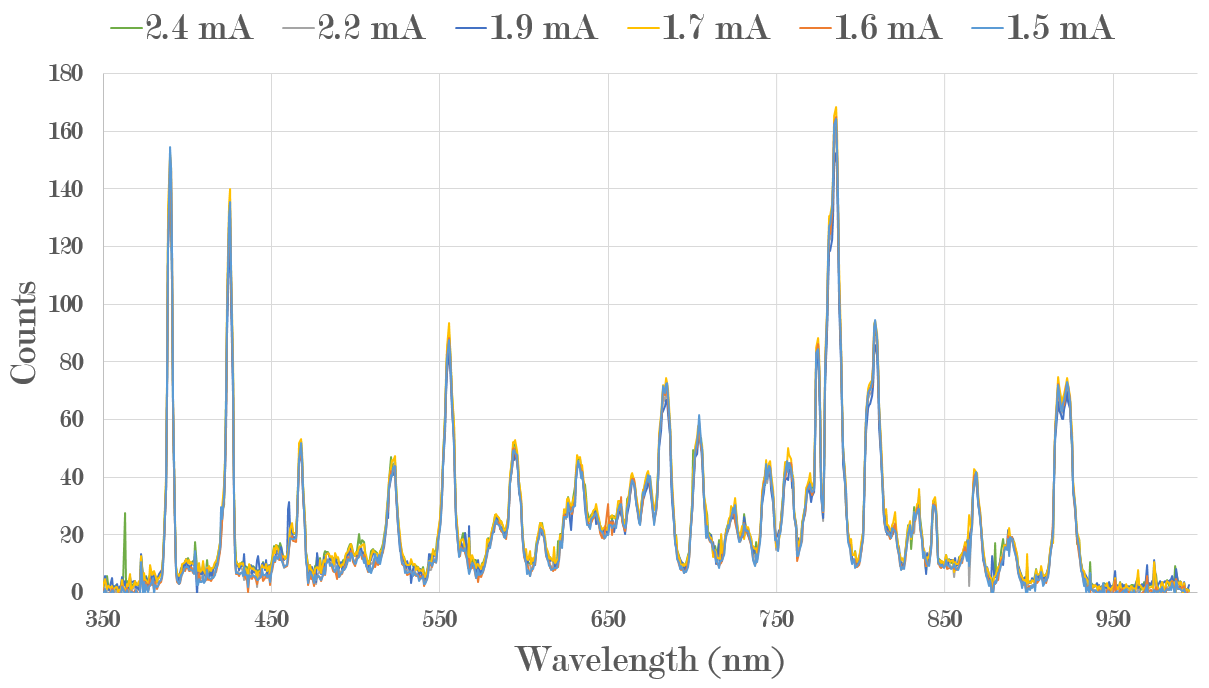}                               
  \caption{The plot shows that the current has no effect on the plasma intensity. The current ranges from 1.5 to 2.4 mA over a 9 day span during the period where the vacuum is most uniform.}
  \label{current}
\end{figure}

The data, illustrated in Figure \ref{current}, supports that the beam has little to no impact on the plasma's spectrum and intensity. During the period from August 5$^{th}$ to the 13$^{th}$, the cyclotron operated with currents between 1.5 mA to 2.4 mA with the four main cavities and the FTC all powered at their beam production values. The same background was subtracted from each time averaged current measurement. Each measurement was a minimum of 1 hour long. Though there was no trend in the current versus the plasma intensity, it was noticed that there is a correlation between the intensity and the date of the measurement, which is explain in the next section.

\subsection{Vacuum}
\begin{figure}[h!]
\centering                                                           
   \includegraphics[scale=.38]{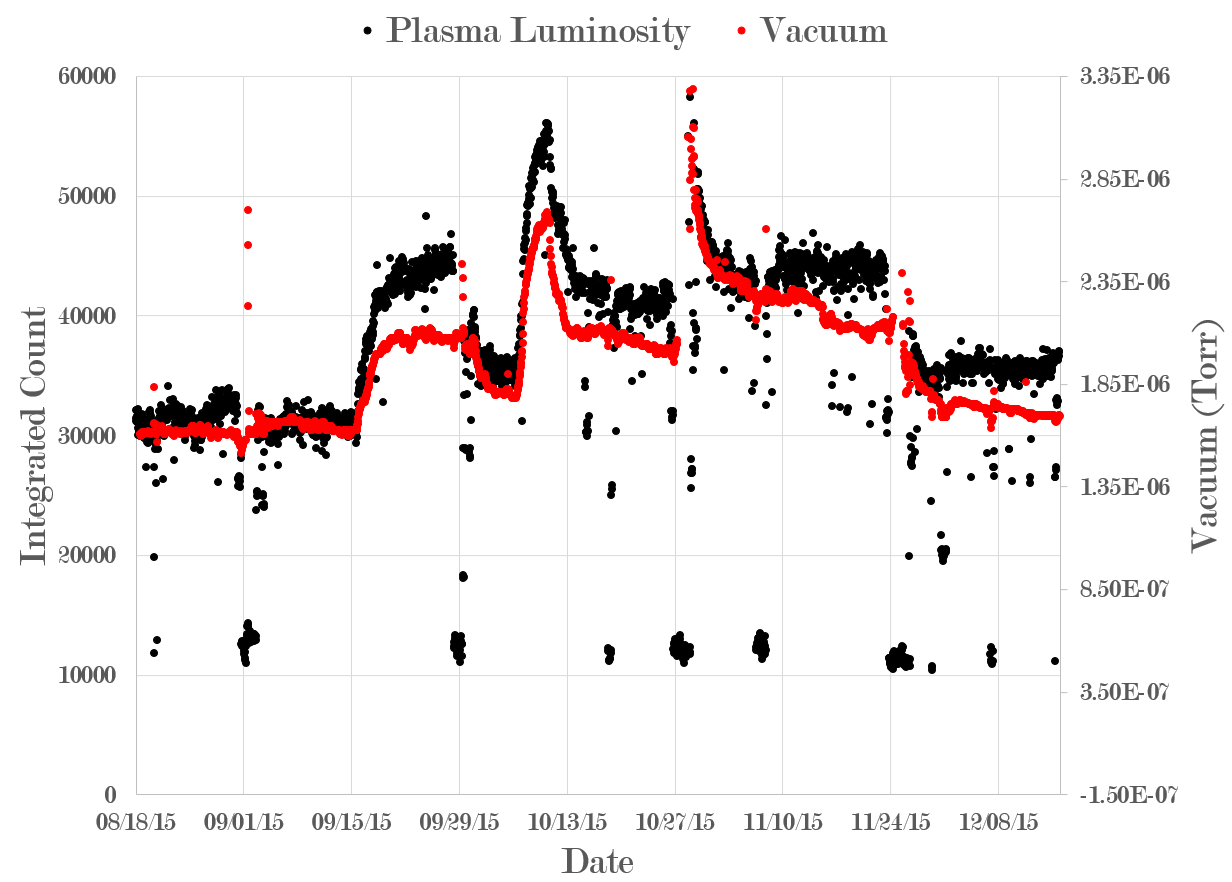}                               
  \caption{The graph is a comparison of the vacuum level and the plasma intensity. The spectrometer takes the total integrated photon count averaged over every hour. The segments where the vacuum level rises and the plasma count falls are maintenance periods. The plasma follows the vacuum closely but always increase over time in comparison to the vacuum level.}
  \label{plasmavac}
\end{figure}

It was initially thought that the plasma was growing rapidly in time. However, when the plasma intensity was compared to the vacuum (thought to be constant) of the cyclotron it became clear that there was a strong dependency. Figure \ref{plasmavac} shows the vacuum and integrated photon count from the spectrometer over the same period of time. The plasma intensity does indeed follow the vacuum, but deviates when the vacuum begins to decrease. Whenever an increase in the vacuum occurs followed by a slow decrease, the plasma count remains at initial level or begins to increase slightly. The window of time in the previous beam current study was kept small and was sampled during the most uniform vacuum level.

   \begin{figure}[h!]
\centering                                                           
   \includegraphics[scale=.34]{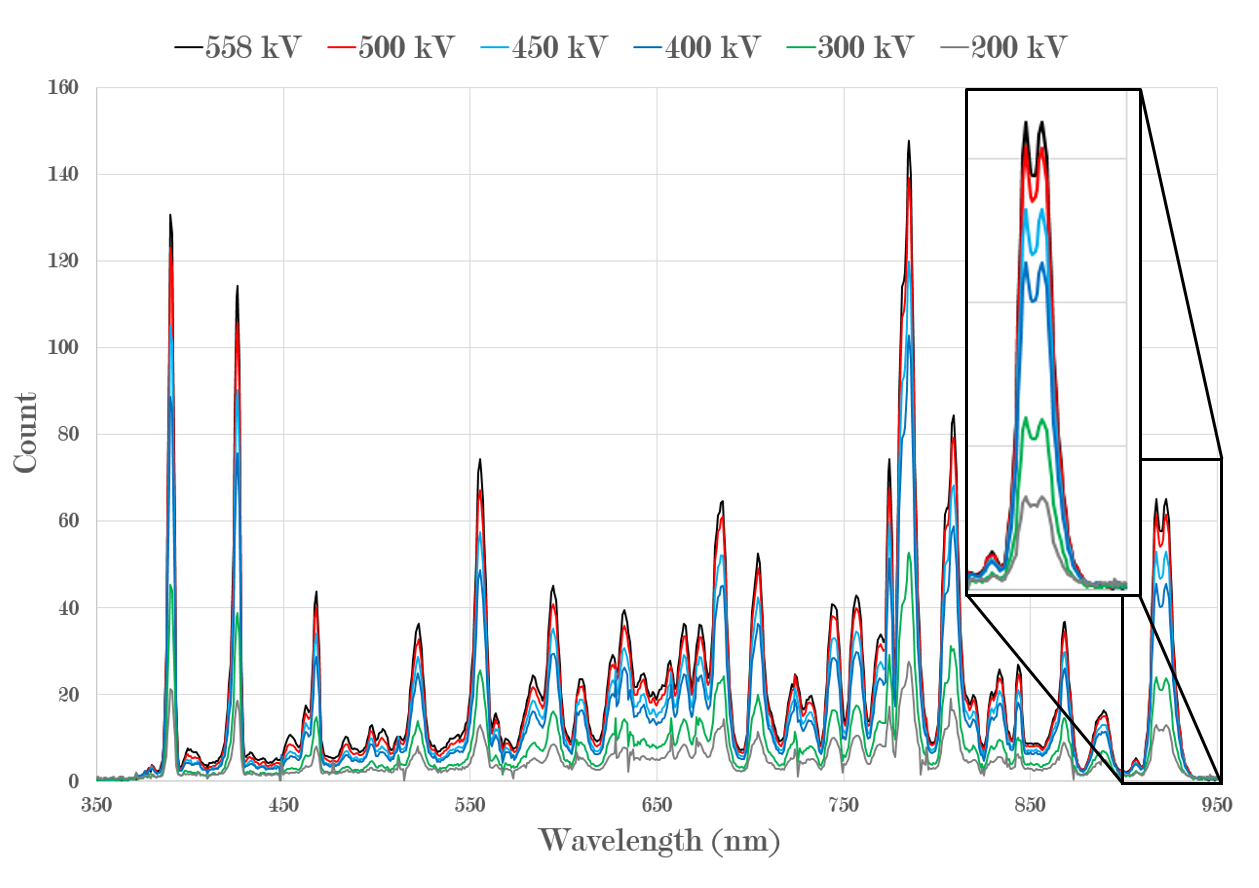}                               
  \caption{RF voltage of the FTC vs. the Plasma Intensity. The graph shows that the voltage in FTC is strongly coupled to the plasma's intensity. The leaking power from FTC is the source of power for the plasma once ignited. Note the plasma is present in the 200 kV and 300 kV measurements even though they are below the 320 kV ignition point.}
  \label{Kav5}
\end{figure} 
 
   \begin{figure}[h!]
\centering                                                           
   \includegraphics[scale=.34]{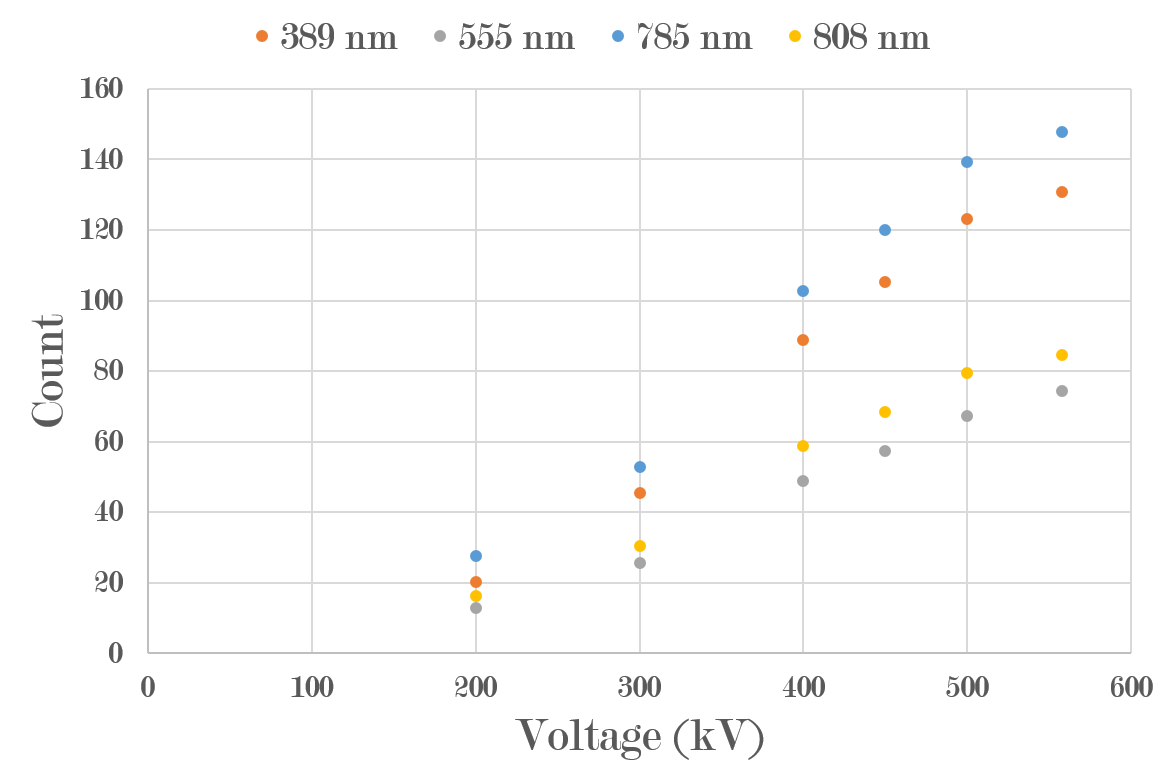}                               
  \caption{This plot illustrates that the plasma count is linearly related to the RF voltage in the cavity.}
  \label{CV}
\end{figure}

\subsection{Flat Top Cavity vs. Plasma Intensity} 
A set of tests were performed to determine the impact of the cavities' voltage on the plasma intensity. All the measurements were taken together over the period of 4 hours such that the background for each measurement would be the same. These experiments verified Goetz's statement that the plasma does not ignite without the FTC \cite{Goetz}. Even with the main cavities fully energized, the plasma would not ignite without the FTC. Figure \ref{Kav5} shows the dependency of the emitted light detected as a function of the voltage applied to FTC only. These measurements were all over 10 minutes and averaged. Isolating a few peaks, a plot showing the intensity versus the voltage was generated. The plasma's intensity is linear with the voltage for the four peaks sampled. The scaling for each peak appears to be consistent as shown in Figure \ref{CV}.

Though there appears to be a linear dependence to the RF voltage, this is not immediately clear as to why this should be the dependence. Light sources in general have their luminosity linear with the power. Perhaps the cause for the linear relationship is that, unlike conventional light sources, the plasma has the capability to grow in volume. Thus one hypothesis is that at low voltages the plasma is in the FTC's neighboring sector magnets, but as power increases it spreads to the other magnets. This can be tested with the insertion of several cameras around the ring. This could explain the a voltage dependence rather than the expected power dependence. Lastly, a plasma's luminosity is highly sensitive to the vacuum as well as the composition and density of the plasma \cite{Plasma2}.

%

The plasma is known to ignite around ~320 kV, thus after this threshold it is possible that the coupling changes in the vacuum space, leading to a non-linear growth of the power leaked out of the cavity with increasing voltage. Additionally, the plasma has been observed well below the ignition point. The determination of the voltage to ignite the plasma and the voltage floor to sustain the plasma will be expounded upon in the next section. However, one can state with confidence that above the ignition voltage the plasma intensity corresponds linearly with the RF voltage.

  \begin{figure}[h!]
\centering                                                           
   \includegraphics[scale=.37]{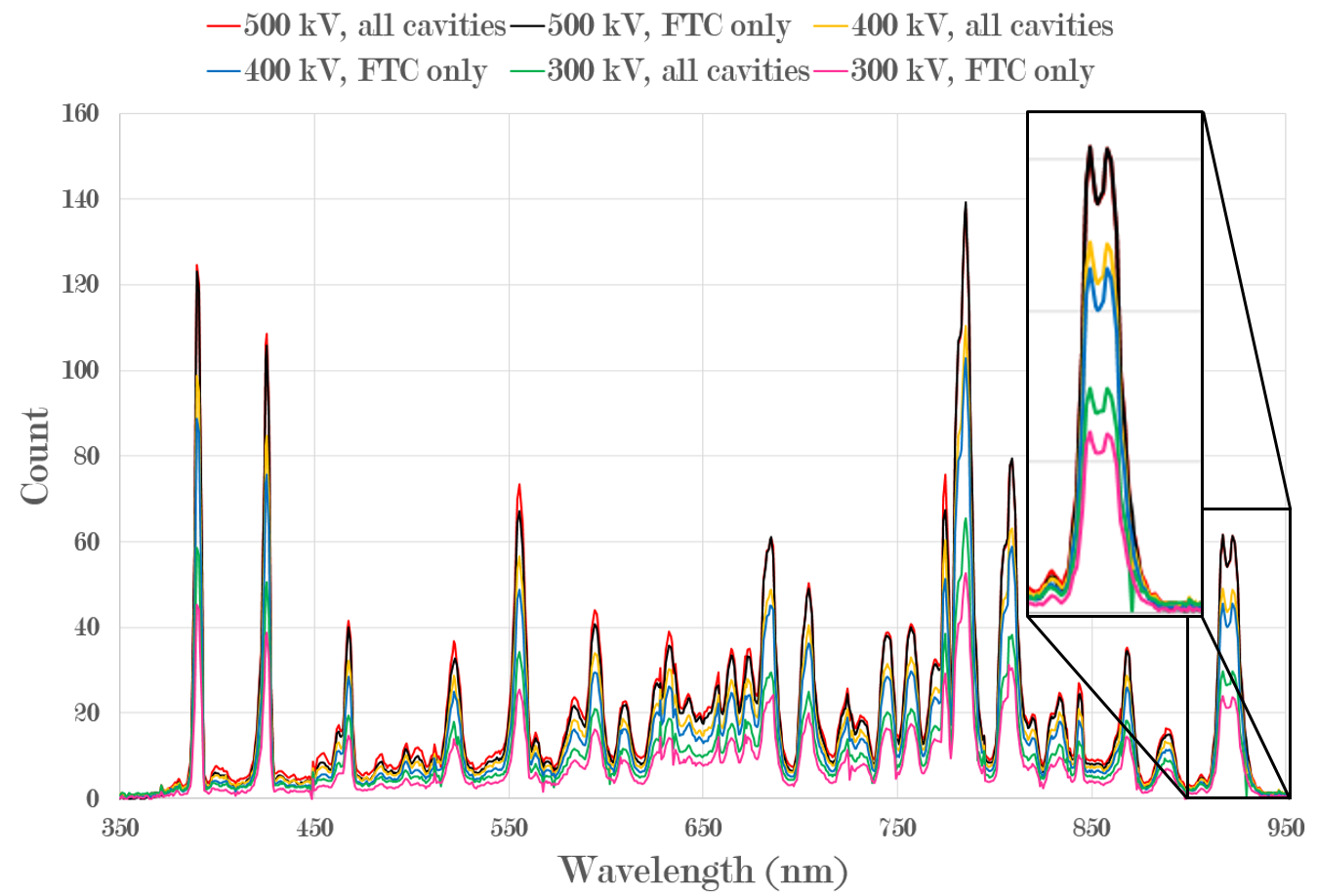}                               
  \caption{The plot shows the spectrum's dependency on the voltage of the main cavities. When the main cavities are energized the spectrum's luminosity does increase. The impact of the 4 main cavities is minimal when higher voltages are applied to the FTC. However their effect becomes more prominent the lower the voltage on the FTC.}
  \label{kav14}
\end{figure}

\subsection{Main Cavities vs. Plasma Intensity} 
In the previous section only the FTC was energized as it was known already that  the plasma could not be ignited without the FTC. To determine the effect of the main cavities, the intensity was measured with the FTC at varying voltages and the main cavities operating at their beam production settings ($\sim$870~kV). Figure \ref{kav14} shows the plasma count for the same voltages with, and without, the main cavities activated. The result is a consistent increase in count when the main cavities are on. The contribution to the count diminishes as the FTC achieves higher voltages. The contribution, or the percent difference between the main cavities being on or off is 1\% when the FTC is at 500 kV, 11\% at 400 kV and 20\% at 300 kV. The reason for this relationship is currently unknown. 

The only difference that is visible in the spectrum during different voltage settings, other than intensity, is that a few additional peaks appear. The cause of the appearance is unknown, but one can conjecture that at higher voltages the signal to noise increases causing the peaks rise above the background noise. Two other hypotheses are that the higher voltage causes a change in energy distribution, or RF field pattern in the cyclotron, and at higher energies other atomic levels are being excited.

\subsection{Summary of Measurements}
The results show that indeed a plasma does exist inside the cyclotron vacuum space. The data confirms that the plasma intensity directly corresponds to the RF voltage on the FTC, as seen in the CCD cameras, and that the beam current does not significantly affect the plasma. The plasma was confirmed not to be ignited when the 4 main cavities were powered exclusively, but the measurements clearly show the plasma intensity is heavily coupled to the vacuum level. In the next section a description of the peaks themselves will be given, highlighting the plasma spectrum is comprised of anticipated materials.

\section{Peak Analysis}

\begin{figure}[h!]
\centering                                                           
   \includegraphics[scale=.34]{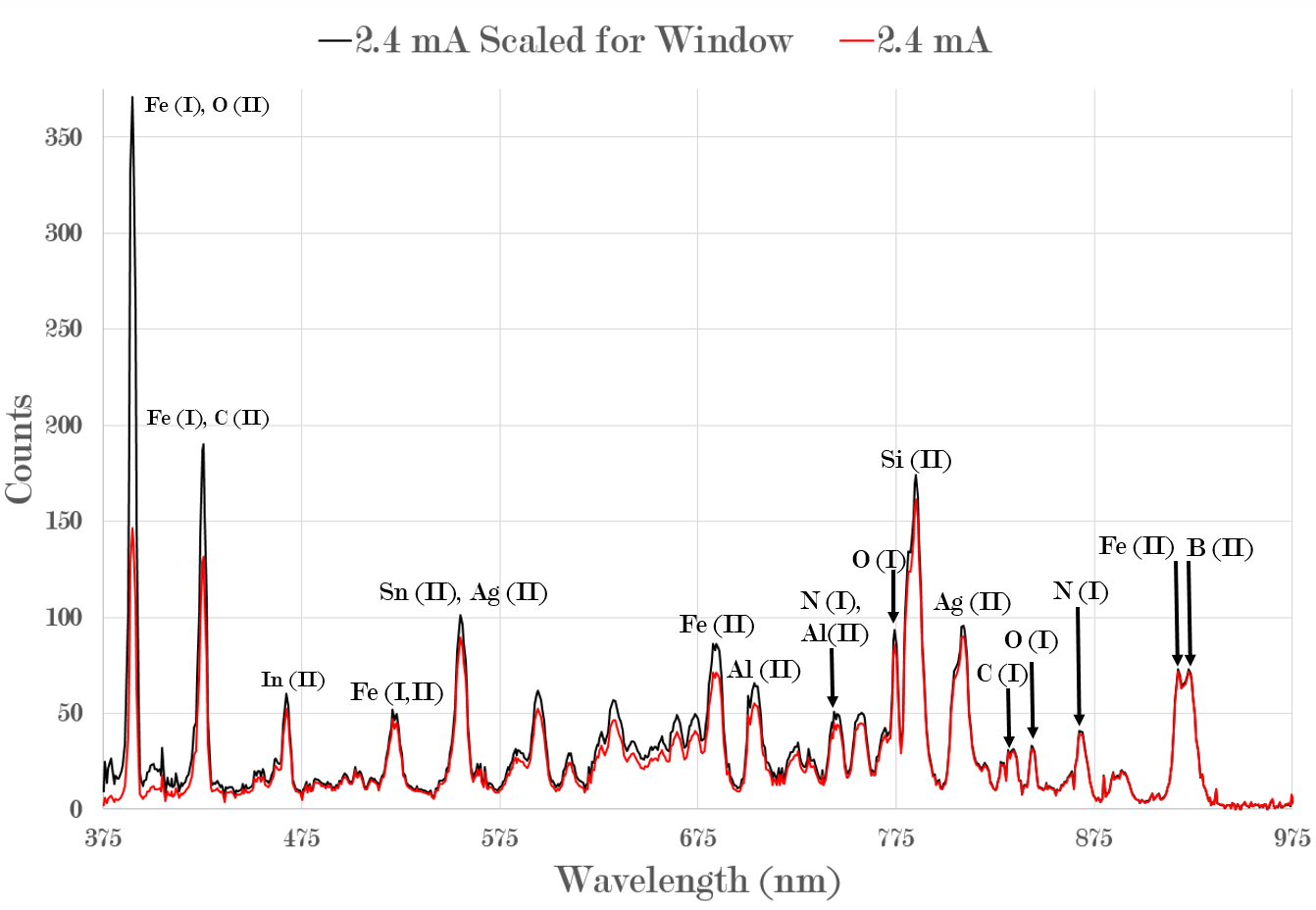}                               
  \caption{The elemental analysis of the spectra using the NIST spectral database for comparison and identification. The raw data (red) is what was read by the spectrometer with the background subtracted. The data scaled for the effect of the coated window is shown in black. This provides a more accurate picture of the intensity of each spectral line.}
  \label{ID}
\end{figure}

To clearly identify the spectral peaks, the spectrum was scaled according to the profile shown in Figure \ref{window-2}. This will allow the most accurate intensity scale between the lines, which might allow for better identification. Figure \ref{ID} shows the unscaled 2.4 mA data and the scaled plot. Additionally the image shows the tentatively identifications of the elements based upon the NIST Atomic Spectra Database \citep{NIST}. Several elements are in mass quantities near the plasma, such as Fe, and are therefore easily understood. However, other elements seen may not come immediately to mind. Indium~(In) and Tin~(Sn) appear due to the ITO coating on the window. Similarly, Boron~(B) and Silicon~(Si) appear due to the borosilicate window. Lastly, Nitrogen~(N) and to a lesser extent, Oxygen~(O) are seen which are from small leaks of the atmosphere into the vacuum space.

Other elements which are less obvious are Carbon (C) and Aluminum (Al). From the previous study performed, the amount of Carbon seen increases substantially. The cause is that a significant part of the interior of the FTC, and the adjacent vacuum chamber, were painted with Aquadag \citep{Aqua}. Aquadag is a graphite solution that dries to form a thin ($\sim$10 $\mu$m), low impedance, layer on a surface to reduce secondary electron emission yield from the surface.  This painting significantly reduced the multipacting located in the cavity and decreased the energy consumption of the cavity because less energy was given to the multipacting electrons.

One signature that was anticipated, but appears to be completely lacking, is the 656 nm or 486 nm Hydrogen line. One would assume there is a substantial water content deposited on the surfaces of the devise during maintenance. However the vacuum teams fills the interior of the vacuum with dry nitrogen rather venting directly to air. This may be a significant reason that the hydrogen lines are not present. 

Identification of the elements is not straight forward as the spectrometer has a 0.77 nm bin size. Due to the plethora of materials inside the vacuum space, several elements can be contributing to an individual peak, or broaden the peak. As a consequence, using the relative intensities for individual elements is challenging. The identifications that are made in Figure \ref{ID} are the most reasonable, probable, in relation to intensity, using the information known about the composition of the vacuum space.

\begin{figure}[h!]
\centering                                                           
   \includegraphics[scale=.39]{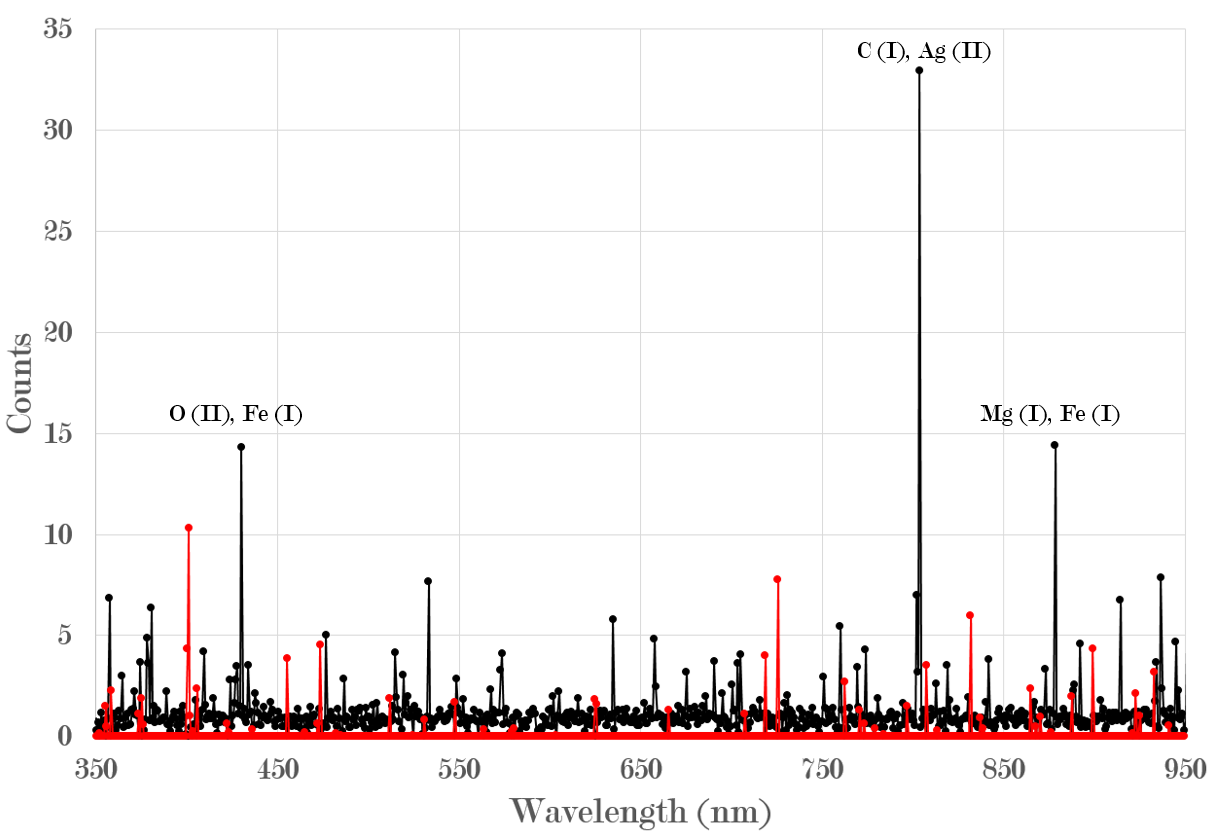}                               
  \caption{This image shows the difference between the spectra with and without one of the main cavities turned on. The two 10 minute measurements were subtracted from each other, and as a result no background has to be taken into account. When the cavity is turned off a distinct set of spectral lines appear. The black peaks are where additional counts were received and the red spikes show where a loss of counts existed. If the spectrometer was calibrated for elements within the vacuum space or had a narrower bin size, the spectrometer could likely be used as a tool to determine what is being heated in the machine. It has been noticed that during the periods when a main cavity is individually off, heating of some components occurs. This is most likely because the field pattern in the cyclotron vacuum chamber is altered.}
  \label{Ag}
\end{figure}

During the voltage varying measurements, one of the main cavities tripped and powered down. This was unnoticed for several measurements. During this period, the ionization monitors increased in count and trim coil temperatures in certain locations increased. Looking at the spectrum, with and without the single main cavity, a substantial increase in certain narrow peaks was noticed. The difference in spectrum when the cavity is on versus off is shown in Figure \ref{Ag}. As a result, if a more sophisticated method of identification that can devised to calibrate the spectrometer, then one could use the device as a means of determining which components are heating under varying RF conditions.

\section{Power Curve Analysis}
A series of measurements were made to provide information on voltage vs. power vs. spectral intensity all on the same day, thus avoiding any uncertainty in the background and vacuum. The power was increased in small steps, and each measurement$'$s duration was 10 minutes. Ionization and vacuum gauges were monitored continuously to determine the ignition point of the plasma and to provide data if other effects are observed. The plasma appeared to ignite near 320 kV, which is lower than the 450 kV needed in 2014, but much better than the 250 kV prior to the Aquadag coating. The power was measured up to $\sim$560~kV, and then dropped again to zero slowly. It was seen that the plasma would extinguish between 20 to 100 kV.

  \begin{figure}[h!]
\centering                                                           
   \includegraphics[scale=.33]{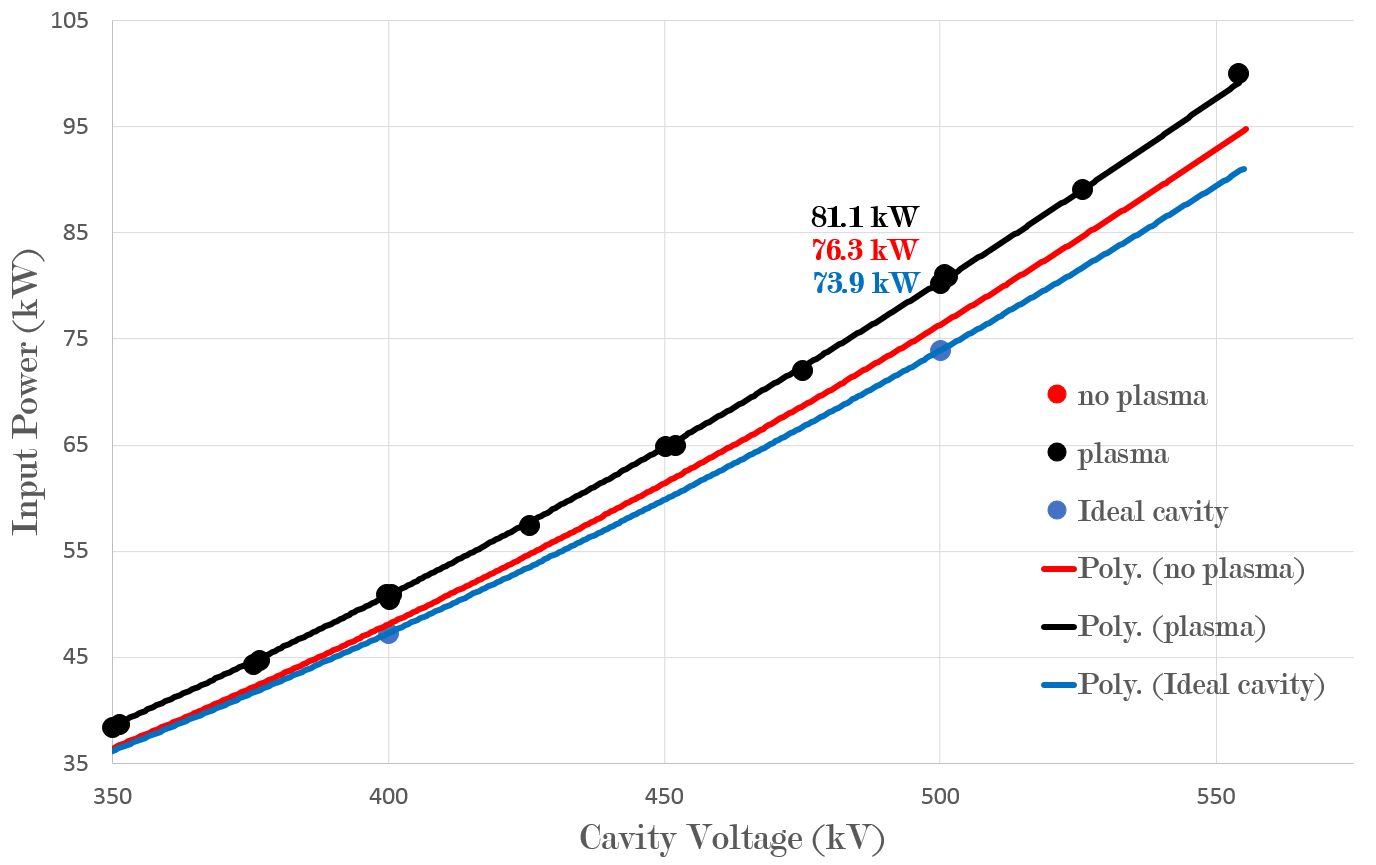}                               
  \caption{The power required by the existing Flat Top cavity (black) to reach a specified voltage is shown, as well as the ideal model projected power requirements (blue). The plasma is ignited around 320 kV, and thus below this threshold several data points were taken. The data was used to create an extrapolation of the power needed in the cavity without the plasma ignited (red). Comparing the plasma vs no plasma data, it can be seen that at 500 kV the plasma pulls 5 kW of power out of the cavity. Comparing the no plasma scenario to the ideal model, one can see a 2.4 kW difference exists, which happens to match the leaked power seen in the finite element model. Image from \cite{cavitypaper}}
  \label{power}
\end{figure}

The data in black, Figure \ref{power}, shows the curve of the data when the plasma is ignited. The data from 0-300 kV, where the plasma was not ignited, was used to create a curve that could be extrapolated to higher voltages. The two curves show that approximately $\sim$5 kW of power at 500 kV is being pulled out of the cavity when the plasma is ignited. 

The blue curve in Figure \ref{power} is generated from a finite element model of the cavity based upon the CAD drawing of the FTC \cite{cavitypaper}. Interferometry measurements were made of the cavities interior and the dimensions of the FTC Finite element model were altered to reflect these findings. The new model, which takes into account the physical measurements, matches the power vs. voltage profile of the cavity when the plasma is not ignited, the red line. The red line shown is an extrapolation of the data taken below 320 kV which was when the plasma ignited. The consequence of these results is that the power leaking out of the FTC due to asymmetry in the cavity is the initial power source of the plasma. Once ignited though, the plasma changes the overall impedance of the cyclotron vacuum chamber and pulls additional power out of the cavity. Overall this provides consistent narrative and directs us to interventions to limit the leaked power from the FTC \cite{cavitypaper}.

\section{Conclusions}
The experiments and the resulting data presented show that the leaking power from the Flat Top cavity is the dominate cause of the plasma. It has been demonstrated that the beam current and the four main cavities have little effect on the plasma. One exception to this statement is that when one or two of the mains are powered down, the resulting field distribution can alter the spectral intensity slightly. If calibrated the spectroscopy system might be a useful tool for determining heating of materials in the cyclotron.

The background measurements, as well as the plasma intensity, grows with time. The reason for this growth is unknown, but has been seen in both 2014 and 2015. The spectra of the plasma contain elements that are all contained within the cyclotron aperture and all elements observed can be accounted for in different elements of the machine. 

The power needed to ignite the plasma is leaked from the Flat Top cavity, but once ignited the plasma draws power out of the cavity causing significantly more power to be pumped into the Flat Top cavity to reach voltage. This is determined from the extrapolation of data when the plasma is not ignited and finite element model based upon interferometry measurements. 

Interventions have been planned to suppress the plasma by attenuating the power leaked out of the cavity and are discussed in Pogue \cite{cavitypaper}. Additionally, elemental analysis using electron backscatter techniques will be performed on components when available to verify the elements being deposited are those being excited in the plasma.

\section{Acknowledgments}
The authors would like to thank Richard Kan for inviting us to look at his camera setup, access to images, and providing insight. The authors would also like to thank the Swiss Plasma Center for fruitful discussions for helping with measurements and providing useful knowledge. Rebekka Bitterli performed preliminary work on the spectroscopy and calibration in the Fall of 2014, and some helpful hints were gleaned from the results as what needed to be done in this (2015) investigation. Lastly, Angelina Parfenova or other for helping access the HIPA data archive and Rasmus Ischebeck for the loan of the spectrometer.

The research leading to these results has received funding from the European Community's Seventh Framework Programme (FP7/2007-2013) under grant agreement n.$^{\circ}$290605 (PSI-FELLOW/COFUND).

\section{Bibliography}
\bibliography{els}

\end{document}